\documentclass[pra,aps,twocolumn,showpacs,epsf]{revtex4}
\usepackage{graphicx}
\usepackage{epsfig}
\usepackage{epsf}
\usepackage{amssymb}
\usepackage{epstopdf}
\usepackage{amsmath}
\usepackage{amstext}
\usepackage{latexsym}
\usepackage[matrix,frame,arrow]{xypic}

\newcommand{\me}{ \mathrm{e}}
\newcommand{\mi}{ \mathrm{i}}

\newcommand{\ket}[1]{| #1 \rangle}
\newcommand{\bra}[1]{\langle #1 |}

\begin{document}

\title{Optimal $1\rightarrow M$ universal quantum cloning via spin networks}
\author{Zhang Jiang$^{1}$}
\author {Qing Chen$^2$}
\author {Shaolong Wan$^{1}$}

\address{
$^{1}$Department of Modern Physics, University of Science and
Technology of China, Hefei 230026, PR China\\
$^{2}$Hefei National Laboratory for Physical Sciences at
Microscale, University of Science and Technology of China, Hefei,
Anhui 230026, PR China
 }
\date{\today}

\begin{abstract}
We present a scheme that transform $1$ qubit to $M$ identical copies
with optimal fidelity via free dynamical evolution of spin star
networks. We show that the Heisenberg XXZ  coupling can fulfill the
challenge. The initial state of the copying machine and the
parameters of the spin Hamiltonian are discussed in detail.
Furthermore we have proposed a feasible method to prepare the
initial state of the copying machine.
\end{abstract}
 \maketitle

One of the most fundamental difference between classical and quantum
information is the no-cloning theorem [1,2]. It states that accurate
cloning of any arbitrary quantum state is impossible. Nevertheless
it doesn't forbid one to clone quantum states approximately. In the
early work of Bu\v{z}ek and Hillery \cite{Buzek}, an optimal
$1\rightarrow2$ universal quantum cloning machine (UQCM) was
proposed where the copying process is input-state independent. And
quantum cloning machine for equatorial qubits which are so called
phase-covariant cloning (PCC) machine was proposed by Bru{\ss}
\emph{et al}. \cite{Bruss2}. Optimal fidelity and optimal quantum
cloning transformation of general $N$ to $M$ ($M>N$) are presented
in [5--8]. It was also shown a UQCM can be realized by a network
consisting of quantum gates \cite{Buzek2}.

Several approaches have been made to realize the unitary
transformations leading to the cloning process
experimentally[10--13]. However, most of these schemes are based on
quantum logic gates and post-selection methods, which need time
modulations. Recently, quantum computation via spin networks based
on Heisenberg couplings was presented [14--27]. One achieve is that
with Heisenberg chains, high fidelity quantum state transfer can be
achieved [15--24]. The most attracting feature of this approach is
that it needn't time modulation for the qubits couplings. Once the
initial states and the evolutional Hamiltonian is determined, the
system can faithfully implement designated computation task through
free dynamical evolution. The whole computational evolution does not
involve any external controlling, which provides relatively longer
decoherence time for the system. Schemes for PCC via spin networks
was proposed in the work of De Chiara \emph{et al}.
\cite{Chiara1,Chiara2}. Chen \emph{et al}. \cite {Chen} further
improved the $1\rightarrow M$ PCC case to an optimal level. However
the optimal UQCM via a spin network is still a challenge.

In this paper, we show that by properly introducing the ancilla
qubies, designing the spin exchange interactions, and choosing the
initial state of the cloning machine, optimal $1\rightarrow M$ UQCM
can be realized via the free evolution of a spin star network
Hamiltonian. Moreover a scheme on preparing the initial state of the
cloning machine have been proposed.

The spin network involved in our scheme forms a star configuration
(See Fig.1(1)). The central qubit (input state) is labeled $I$, the
$M$ target qubits labeled $T$, and the $M-2$ ancillas labeled $A$.
We start with the conventional Heisenberg XXZ  coupling Hamiltonian
without an externally applied magnetic field.
\begin{eqnarray}
H&=&\frac{\mathcal{J}_1}{2}\sum_{i=1}^{M}(\sigma_I^x\sigma_{T_i}^x+\sigma_I^y\sigma_{T_i}^y+
\lambda_1\:\sigma_I^z\sigma_{T_i}^z)\nonumber\\
&+&\frac{\mathcal{J}_2}{2}\sum_{i=1}^{M-2}(\sigma_I^x\sigma_{A_i}^x+\sigma_I^y\sigma_{A_i}^y+
\lambda_2\:\sigma_I^z\sigma_{A_i}^z),\label{H}
\end{eqnarray}
where $\sigma^{x,y,z}_I$, $\sigma^{x,y,z}_{T_i}$,
$\sigma^{x,y,z}_{A_i}$ are Pauli matrices of the input particle, the
target qubit, and the ancilla qubits respectively(we introduce $M-2$
ancilla qubits), $\mathcal{J}_1$ and $\mathcal{J}_2$ are the
exchange spin coupling coefficients between the input qubit with the
target qubits and the ancilla qubits respectively, $\lambda_1$ and
$\lambda_2$ are the anisotropy parameters (when $\lambda=0$, the
Hamiltonian reduces to $XX$ model while $\lambda=1$ it corresponds
to Heisenberg model).

Following with Gisin and Massar \cite{Gisin2} we suppose the unitary
transformation for optimal $1\rightarrow M$ cloning take the form:
\begin{eqnarray}
U_{1,M}\ket{\uparrow}_I\otimes
\ket{R}&=&\sum_{i=0}^{M-1}\gamma_i\:\ket{S(M,M-i)}_T\otimes \ket{R_i},\label{t1}\\
U_{1,M}\ket{\downarrow}_I\otimes
\ket{R}&=&\sum_{i=0}^{M-1}\gamma_{M-1-i}\nonumber\\
&&\times\ket{S(M,M-1-i))}_T\otimes \ket{R_i},\label{t2}\\
\gamma_i&=&\sqrt{\frac{2(M-i)}{M(M+1)}},\nonumber
\end{eqnarray}
where $U_{1,M}=\me^{-\mi H t_0}$ ($t_0$ is the evolution time)
denotes the free evolution of the spin system, $\ket{R}$ denotes the
initial state of the copying machine and $M$ blank copies.
$\ket{S(M,i)}_T$ is the normalized symmetry state of the $M$ target
qubits with $i$ spins up. $\ket{R_i}$ are orthogonal normalized
sates of the ancilla qubies (here include the input qubit). We
choose the initial state $\ket{R}$ as follows:
\begin{eqnarray}
\ket{R}&=&C\sum_{i=1}^{M-1}\sqrt{i(M-i)}\;\ket{a_i},\label{R}\\
\ket{a_i}&=&\ket{S(M,i)}_T\otimes\ket{S(M-2,M-1-i)}_A,\nonumber
\end{eqnarray}
where $C=\sqrt{\frac{6}{(M-1)M(M+1)}}$ is the normalization factor,
$\ket{S(M-2,M-1-i)}_A$ is the normalized symmetry state of the $M-2$
ancilla qubits. Noticing $\ket{R}$ is invariant under the spin
flipping operation, we will show later by spin flipping both sides
of Eq.(\ref{t1}), Eq.(\ref{t2}) is automatically satisfied.

\begin{figure}
\hspace{-2em}\epsfig{file=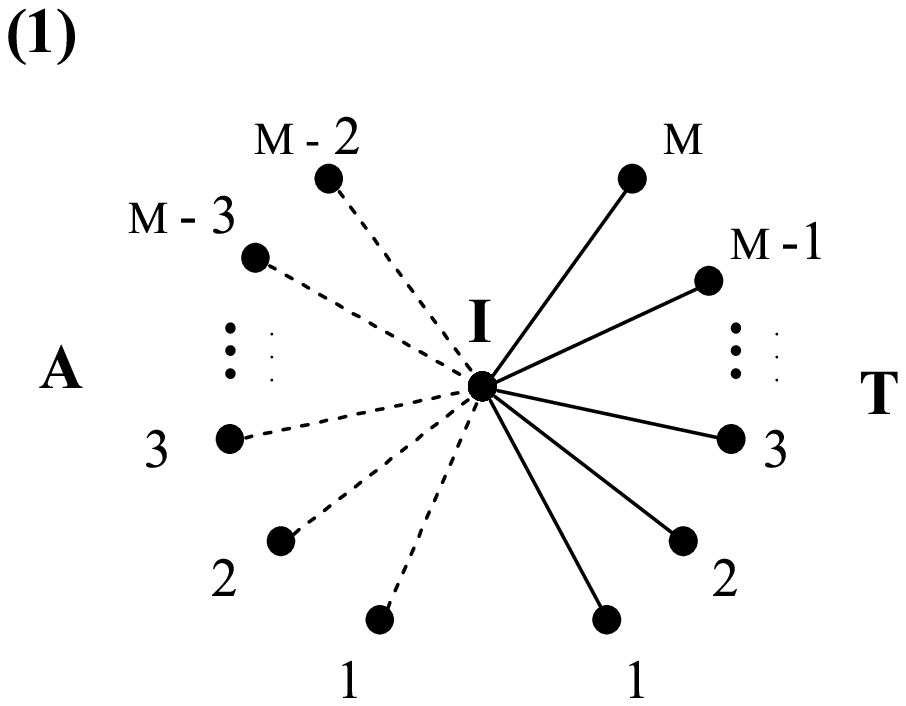,height=4cm}\epsfig{file=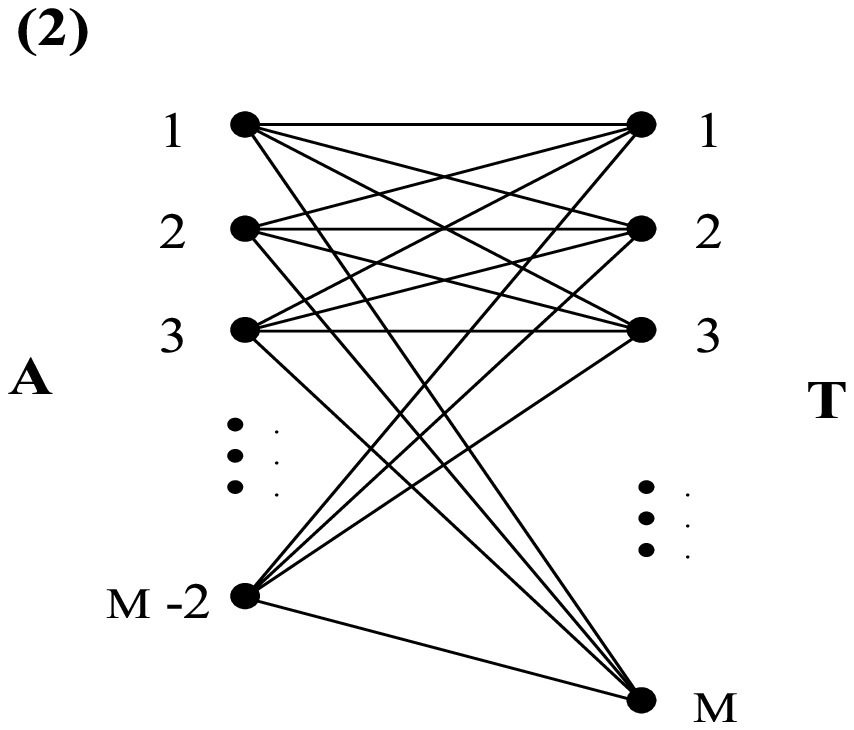,height=4cm}
\caption{ (1) Spin star network for $1\rightarrow M$ UQCM. The left
side spins form the ancilla, and spins on the right side form the
target particles. (2) Spin network employed for generating initial
states. Each ancilla spin interacts with all the target spins.}
\end{figure}

First we discuss the conditions to satisfy Eq.(\ref{t1}). Instead of
studying all the states in the Hilbert space of the Hailtonian
(\ref{H}), we would rather to introduce a two dimensional subspace
$\mathcal{H}_{ab}$ (we use $\psi_{ab}$ to note states in this
subspace, $\psi_{ab}^\bot$ to note states orthogonal to this
subspace), which is spanned by two basic normalized orthogonal
states $\ket{a}$, and $\ket{b}$
\begin{eqnarray}
&&\ket{a}=\ket{\uparrow}_I\otimes\ket{R}\;, \label{a}\\
&&\ket{b}=\frac{\sqrt{2}C}{2}\ket{\downarrow}_I\otimes\big(\sum_{j=1}^{M-1}\sqrt{j(j+1)}\:\ket{b_j}\big)\;,\label{b}\\
&&\ket{b_j}=\ket{S(M,j+1)}_T\otimes\ket{S(M-2,M-1-j)}_A\;.\nonumber
\end{eqnarray}
Notice $\ket{a}$ is our initial state for Eq.(\ref{t1}), and we will
show that some linear combination of these two states has the same
form of the righthand side of Eq.(\ref{t1}). We find if the
parameters of the spin Hamiltonian (\ref{H}) obey the following
relations
\begin{equation}\mathcal{J}_1=-\mathcal{J}_2=\mathcal{J},\;\;\;\lambda_1=-\lambda_2=\lambda.\label{co}
\end{equation}
the subspace we choosing is closed, i.e.,
$\bra{\psi_{ab}}H\ket{\psi_{ab}^\bot}=0$. Then it is convenient for
us to calculate the free evolution of the system in this two
dimensional subspace. It is useful to rewrite the Hamiltonian
(\ref{H}) with the Ladder operators. Using the relations (\ref{co})
the Hamiltonian take the form:
\begin{eqnarray}
&&\hspace{-2em}s^{\pm}_I=(\sigma^x_I\pm
i\sigma^y_I)/2,\;\;J^{\pm}_T=
\sum_{T} s^{\pm}_{T},\;\;J^{\pm}_A=\sum_{A} s^{\pm}_{A},\nonumber\\
&&\hspace{-2em}s^z_I=\sigma^z_0/2,\;\;
J^z_T=\sum_T\sigma^{z}_T/2,\;\;J^z_A=\sum_A\sigma^{z}_A/2,\nonumber\\
H&=&\mathcal{J}\big(s^+_I (J^-_T-J^-_A)+s^-_I
(J^+_T-J^+_A)+2\lambda\:s^z_I (J^z_T+J^z_A)\big).\nonumber
\end{eqnarray}
With this representation of the spin Hamiltonian it is easy for us
to calculate $H$ act upon our bases.
\begin{eqnarray}
H\ket{a}&=&\sqrt{2}\mathcal{J}\ket{b},\nonumber\\
H\ket{b}&=&-\mathcal{J}\lambda\ket{b}+\sqrt{2}\mathcal{J}\ket{a}\nonumber.
\end{eqnarray}
Thus, the Hamiltonian is closed in the subspace $\mathcal{H}_{ab}$,
and we can write the matrix form of $H$ in $\mathcal{H}_{ab}$,
\begin{equation}
\tilde{H}=\begin{pmatrix}
  0 & \sqrt{2}\mathcal{J} \\
  \sqrt{2}\mathcal{J} & -\mathcal{J}\lambda
\end{pmatrix}\nonumber.
\end{equation}
This is the key point of our scheme. Despite how large $M$ is, such
a two dimensional space always exist as long as the condition
(\ref{co}) is kept. Now, our problem reduced to a two dimensional
quantum evolution in $\mathcal{H}_{ab}$, the unitary transformation
$U_{1,M}(t)$ takes the form:
\begin{eqnarray}
\tilde{U}_{1,M}(t)&=&\exp\{-\mi\tilde{H}t\}\nonumber\\
&=&\me^{\mi\mathcal{J}t\lambda/2}\big(\cos(\frac{1}{2}\mathcal{J}t\sqrt{\lambda^2+8})\:I\nonumber\\
&&-\mi\frac{\sin(\frac{1}
{2}\mathcal{J}t\sqrt{\lambda^2+8})}{\sqrt{\lambda^2+8}}\:(\lambda\sigma_z+2\sqrt{2}\sigma_x)\big),\nonumber
\end{eqnarray}
where $\sigma_z$ and $\sigma_x$ are pauli matrices in
$\mathcal{H}_{ab}$. We choose the anisotropy parameter
\begin{equation}\lambda=2.\label{anisotropy}\end{equation}
As our initial state for Eq.(\ref{t1}) is $\ket{a}$, after having
evolved for $t$,
\begin{eqnarray}
\ket{a(t)}&=&\tilde{U}_{1,M}(t)\ket{a}\nonumber\\
&=&\me^{\mi\mathcal{J}t}\big(\:(\cos\varphi-\frac{\mi\sin\varphi}{\sqrt{3}})\ket{a}-\frac{\mi\sqrt{2}\sin\varphi}{\sqrt{3}}\ket{b}\:\big),\nonumber
\end{eqnarray}
where $\varphi=\sqrt{3}\mathcal{J}t$ is the rescaled time
parameter. When $\varphi=\pi/2$, i.e,
\begin{equation}
t=t_0=\frac{\sqrt{3}\pi}{6\mathcal{J}},\label{t}
\end{equation}
the state of the system take the following form,
\begin{eqnarray}
\ket{a(t_0)}&=&-\mi\me^{\mi\mathcal{J}t}(\sqrt{\frac{1}{3}}\;\ket{a}+\sqrt{\frac{2}{3}}\;\ket{b})\nonumber\\
&=&-\mi\me^{\mi\mathcal{J}t}\sum_{i=0}^{M-1}\gamma_i\:\ket{S(M,M-i)}_T\nonumber\\
&&\hspace{-2em}\otimes\;\ket{S(M-1,i)}_{A\otimes I}\nonumber
\end{eqnarray}
where $\ket{S(M-1,i)}_{A\otimes I}$ denotes the normalized
symmetry state in the direct product space of the input qubit and
ancilla qubits. The state $\ket{a(t_0)}$ is exact the same form as
the righthand side of Eq.(\ref{t1}). The orthogonal normalized
states $\ket{R_i}$ take the form:
\begin{equation}
\ket{R_i}=-\mi\me^{\mi\mathcal{J}t}\ket{S(M-1,i)}_{A\otimes
I}\nonumber.
\end{equation}
To go further, we introduce the spin flipping operator,
$$P=P^{-1}=\sigma_I^x\:(\prod_i^M\sigma_{T_i}^x)\:(\prod_j^{M-2}\sigma_{A_j}^x).$$
This unitary operation flip all the spins in our consideration. It
is easy to see that the Heisenberg XXZ  spin Hamiltonian (\ref{H})
is invariant under such operation, i.e., $PHP^{-1}=H$. The initial
state for Eq.(\ref{t2}) is
$\ket{\downarrow}_I\otimes\ket{R}=P\;\ket{a}=\ket{a_{\scriptscriptstyle
P}}$, after having evolved for $t$
$$\ket{a_{\scriptscriptstyle P}(t)}=U_{1,M}(t)\ket{a_{\scriptscriptstyle P}}=P\;\tilde{U}_{1,M}(t)\ket{a}=P\;\ket{a(t)}.$$
When the evolution time $t=t_0$,
\begin{eqnarray}
\ket{a_{\scriptscriptstyle P}(t_0)}&=&P\;\ket{a(t_0)}
=-\mi\me^{\mi\mathcal{J}t}\sum_{i=0}^{M-1}\gamma_i\:\ket{S(M,i)}_T\nonumber\\
&&\hspace{-2em}\otimes\;\ket{S(M-1,M-1-i)}_{A\otimes I}.\nonumber
\end{eqnarray}
It is exact the same form as the righthand side of Eq.(\ref{t2}).
The above calculation show that we can find such conditions
(\ref{R},\ref{co},\ref{anisotropy},\ref{t}) satisfying Eq.(\ref{t1},
\ref{t2}) simultaneously, i.e, the optimal cloning can be fulfilled
under such conditions.

\begin{figure}
\hspace{-2em}\epsfig{file=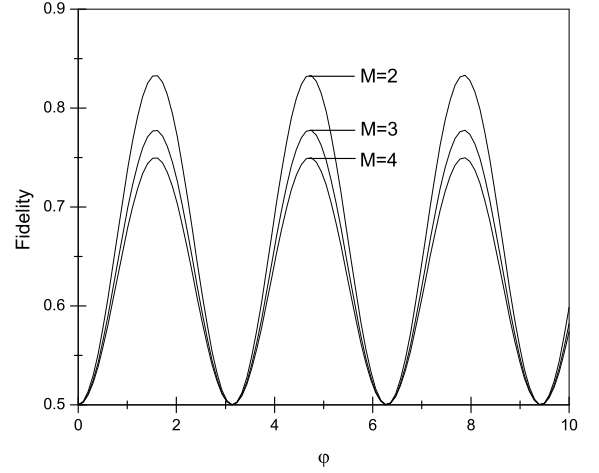,height=8cm} \caption{The
input state independent fidelity of a single copy as a function of
the rescaled time $\varphi$ ($\varphi=\sqrt{3}\mathcal{J}t$), for
$M=2,3,4$. When $\varphi=n\pi$ the fidelity equals 1/2, and when
$\varphi=(2n+1)\pi/2$ the fidelity reaches it's optimal bound
$(2M+1)/3M$.}
\end{figure}
One interesting thing is that through the beginning to the end of
this free evolution the fidelity of a single copy to the input is
independent of the input state (a universal cloning). Suppose the
input state is:
$\ket{\mathrm{input}}_I=\alpha\ket{\uparrow}_I+\beta\ket{\downarrow}_I.$
After having evolved for $t$, the state of the system take the form:
$\ket{t}=\alpha\ket{a(t)}+\beta\ket{a_{\scriptscriptstyle P}(t)}$.
The reduced density matrix of a single copy at $t$ can be calculated
directly,
\begin{eqnarray}
\hspace{-1.5em}\rho &=&\frac{\cos^2\!\varphi}{2}\begin{pmatrix}
  1 & 0 \\
  0 & 1
\end{pmatrix}
+\frac{\sin^2\!\varphi}{3M}\times\nonumber\\
&&\hspace{-1.5em}\begin{small}\begin{pmatrix}
  \alpha^2(1+2M)+\beta^2(M-1) & \beta^*\alpha(M+2) \\
  \alpha^*\beta(M+2) & \beta^2(1+2M)+\alpha^2(M-1)
\end{pmatrix}.\nonumber\end{small}
\end{eqnarray}
The fidelity of this copy is
\begin{equation}
F=\frac{1}{2}\cos^2\!\varphi+\frac{2M+1}{3M}\sin^2\!\varphi.\label{F}
\end{equation}
$F$ is only a function of the rescaled time $\varphi$
($\varphi=\sqrt{3}\mathcal{J}t$). So the whole cloning process is
input state independent. When $t=0$ the fidelity is $1/2$, and when
$t=t_0$ the fidelity reaches it's optimal bound $(2M+1)/3M$ (see
Fig.2).

One shortcoming of quantum cloning based on logic gates is the
circuit becomes more complicated as $M$ increases. As a result, when
$M$ is large it may be difficult for one to go through the copying
process before the state having been decoherenced. However, the
evolution time of our scheme is $t_0=\sqrt{3}\pi/(12\mathcal{J})$,
which is independent of $M$. This is an advantage to fulfill large
$M$ cloning.

The problem now is how to prepare the initial state  (\ref{R}). For
$M=2$ ($M=3$), $\ket{R}$ is two (four) particle symmetry state. But
for $M>3$, $\ket{R}$ is not simply a symmetry state. Interestingly,
we find that $\ket{R}$ is exactly the ground state of some spin
Hamiltonian. And it is feasible for one to prepare it by just
cooling the system. Such Hamiltonian is consisted of two parts
\begin{equation}
H'=H'_0+H'_1.\label{H'}
\end{equation}
$H'_0$ is the part with Heisenberg XXZ  coupling ($\lambda=-1$)
between the target qubits and the ancilla qubits (Fig.1(2)),
\begin{equation}
H'_0=\mathcal{J}'(J^+_T J^-_A+J^-_T J^+_A-2J^z_T J^z_A),\label{H'0}
\end{equation}
where $\mathcal{J}'$ is the spin coupling coefficient, $J_T$ and
$J_A$ are total angular momentum operators of the target qubits and
the ancilla qubits respectively. $H'_1$ is the part with Ising
coupling between all the qubits,
\begin{eqnarray}
H'_1&=&\frac{\Delta}{2}\big(\sum_{\substack{i=1\\k< i}}^M
\sigma^z_{T_i} \sigma^z_{T_k} +\sum_{\substack{j=1\\k<
j}}^{M-2}\sigma^z_{A_j}
\sigma^z_{A_k}+\sum_{i=1}^M\sum_{j=1}^{M-2}\sigma^z_{T_i}
\sigma^z_{A_j}\big)\nonumber\\
&=&\Delta(J^z)^2-\frac{\Delta(M-1)}{2},\label{H'1}
\end{eqnarray}
where $\Delta$ is the coupling coefficient. These two parts are
commute, $[H'_0 , H'_1]=0$. We find $\ket{R}$ is an eigenvector of
$H'_0$ and $H'_1$ simultaneously
\begin{eqnarray}
H'_0\ket{R}=\frac{\mathcal{J}'(M^2-4)}{2}\;\ket{R},\nonumber\\
H'_1\ket{R}=-\frac{\Delta(M-1)}{2}\;\ket{R}.\nonumber
\end{eqnarray}
To prove $\ket{R}$ is the ground state we solve the spectrum of
$H'$. We introduce the unitary operator $Q_T$ to act on the target
qubits (it is equivalence to introduce $Q_A$ acting on the ancilla
qubits),
$$Q_T=Q^{-1}_T=\prod_{i=1}^M\sigma^z_{T_i}.$$
This unitary operation transforms $H'_0$ to the Heisenberg
Hamiltonian and leaves $H'_1$ unchange,
\begin{eqnarray}
Q_TH'_0Q_T^{-1}&=&-\mathcal{J}'(J^+_T J^-_A+J^-_T J^+_A+2J^z_T
J^z_A)\nonumber\\
Q_TH'_1Q_T^{-1}&=&H'_1\nonumber
\end{eqnarray}
So the spectrum of $H'$ is
\begin{eqnarray}
E'&=&-\mathcal{J}'\big(j(j+1)-j_T(j_T+1)-j_A(j_A+1)\big)\nonumber\\
&&+\Delta(j^z)^2-\frac{\Delta(M-1)}{2}\;,\nonumber
\end{eqnarray}
where $j$ ($j^z$) is the total ($z$ component) angular momentum
quantum number of the transformed Hamiltonian. If we choose
$\mathcal{J}'<0$, and $\Delta >0$, the nondegenerate ground state
energy of $H'$ is $\mathcal{J}'(M^2-4)/2-\Delta(M-1)/2$ ($j=1,
j_T=M/2, j_A=(M-2)/2, j^z=0$), which is just the eigenvalue of
$\ket{R}$. So far, we have proved $\ket{R}$ is the ground state of
$H'$. Thus the initial state of the copying machine can be
prepared by cooling the system. No measurement is involved in this
implementation, and also we needn't any time modulation of the
Hamiltonian.

Through out this paper, optimal UQCM that produce $M$ copies out of
a single input via free evolution of spin star networks has been
discussed. We have proved for arbitrary $M$ the unitary evolution
can be fulfilled in a two dimensional subspace. Using this character
we find the analytical solutions for the optimal $1\rightarrow M$
universal cloning process. Through this process the fidelity keeps
input state independent, and it reaches the optimal bound at
$t=\sqrt{3}\pi/6\mathcal{J}$, which is independent of $M$. Also we
have studied the initial state of the coping machine in detail, and
find it is exactly the ground state of some spin Hamiltonian (only
quadratic terms are involved). Thus, the preparation of the initial
state can be accomplished by cooling such systems. No measurement
and time modulation is involved here. Therefore our result opens up
a promising prospect towards robust optimal UQCM. Such a prospect is
relevant for several experimental systems \cite{{Romito},{Peng}}.

\end{document}